\documentclass[english,12pt,a4paper,twoside]{article}
\usepackage{babel,epsf}
\usepackage[latin1]{inputenc}
\usepackage[T1]{fontenc}
\usepackage[dvips]{graphics,color}
\usepackage{latexsym}
\usepackage{makeidx}
\usepackage{graphicx}
\usepackage{dcolumn}
\usepackage{amsmath}
\usepackage{epsfig}
\usepackage{amsfonts}
\usepackage{amstext}
\usepackage{amssymb}
\usepackage{revsymb}
\usepackage{mathrsfs}
\begin{document}

\title{On what does not expand in an expanding universe: a very simple model}
\author{Nivaldo A. Lemos\\
{\small{Departamento de F\'{\i}sica}}\\
{\small{Universidade Federal Fluminense}}\\
{\small{Av. Litor\^anea s/n, Boa Viagem - CEP 24210-340}}\\
{\small{Niter\'oi - Rio de Janeiro}}\\
{\small{Brazil}}\\
E-mail: nivaldo@if.uff.br }
\maketitle

\begin{abstract}
As the separation between galaxies increases owing to the expansion of the universe, galaxies themselves and smaller bound structures do not grow. An accurate description of the dynamics of cosmic structures requires the full apparatus of general relativity.  In order to gain a fairly satisfactory understanding of what does not expand in an expanding universe, however, it suffices to take the harmonic oscillator as prototype of a bound system. More precisely, we show that a study of the quantum dynamics of a nonrelativistic harmonic oscillator in an expanding universe makes it clear that most bound systems do not take part in the overall cosmic expansion. The  analysis is elementary and indicates
that whether a bound structure partakes in the expansion partially or  not at all is essentially determined by a characteristic time scale associated with it.

\end{abstract}

\newpage

\section{Introduction}

The expansion of the universe is viewed as a stretching of space that increases the separation between galaxies, but the galaxies themselves and smaller bound structures do not expand.  After all, if everything expanded at the same rate the expansion itself would not be observable.  As discussed by Anderson \cite{Anderson}, cosmological theory alone does not seem to give a clear-cut answer as to the scale at which bound systems start to feel the influence of the general expansion, a point also emphasized by Bonnor \cite{Bonnor}, who  provides a short history of the issue. Regarding structures larger than galaxies, the possible expansion of galaxy clusters was considered by  Noerdlinger and Petrosian \cite{Noerdlinger}.

In a recent work, Price and Romano \cite{Price} discussed in an elementary way the behavior of a classical atom in an expanding universe and concluded that, as expected,  atoms do not partake in the cosmological expansion. This agrees with more advanced and detailed analyses, which also show that a classical hydrogen atom either does not grow at all  or does grow but at a rate negligible compared with the general cosmic expansion \cite{Bonnor}.
But one should not fail to notice that these investigations are open to the criticism that atoms should be treated by quantum mechanics. 
However, to our knowledge,  quantum mechanical treatments of a hydrogen atom in cosmological spacetimes  have  been concerned not with the  atom's   possible stretching, but  with energy-level shifts as well as whether the atom,   taken as a clock,   shows proper  time \cite{Audretsch}.

Certainly a complete description of what does not expand in an expanding universe requires the full apparatus of general relativity. Our aim here is much less ambitious: it is our intention, in line with investigations such as those of Bonnor \cite{Bonnor} or Price and Romano \cite{Price}, to formulate a model as elementary  as possible of how and to what degree the overall cosmic expansion affects bound structures.

 The harmonic oscillator is a  bound system simpler than the hydrogen atom. This simplicity brings about the important advantage that not only the classical dynamics but also  the quantum dynamics of a harmonic oscillator in an expanding universe is exactly soluble. This allows an exact discussion of to what extent the oscillator is influenced by the cosmic expansion.  Although the harmonic oscillator is certainly  a  crude model, it does acccount for uniform circular motion as well as some elliptic motions, which are superpositions of two perpendicular harmonic oscillations. Furthermore, its quantized version holds not only for small but also for large systems, since classical mechanics is a limiting case of quantum mechanics.  By taking a harmonic oscillator as the archetypical bound system, we find, in agreement with Anderson \cite{Anderson}, that ``all physical systems, big or small, feel the effect of the cosmic expansion in one way or another.'' Our approach suggests that the  decisive factor is a  characteristic time scale associated with the system.

\section{Model and Results}

As argued in \cite{Price}, nonrelativistic dynamics is adequate for the description of the relevant bound structures in an expanding universe. It has been shown by Lemos and Natividade \cite{Lemos} that the Lagrangian for a nonrelativistic particle in a spatially-flat Friedmann-Robertson-Walker universe is  
\begin{equation}
\label{lagrangian}
L = \frac{mv^2}{2} + \frac{m{\ddot a}}{2a}r^2 - V({\bf r})
\end{equation}
where $a(t)$ is the scale factor, $t$ is cosmic time, ${\bf r}=(x^1,x^2, x^3)$ is the position vector whose components are the physical position coordinates of the particle, $v^2= \vert {\dot{\bf r}}\vert^2$   and $V({\bf r})$ is the potential energy associated with  forces other than gravity. The second term on the right-hand side of Eq. (\ref{lagrangian}) is responsible for the radial acceleration given by equation (4) in Price and Romano \cite{Price}.

With the center of force at the origin, the potential for an isotropic oscillator is $V({\bf r}) = m\omega_c^2r^2/2$. The Lagrangian (\ref{lagrangian}) becomes
\begin{equation}
\label{lagrangian-time-dependent-frequency}
L = \frac{mv^2}{2} - \frac{m}{2} \omega^2(t)r^2 \, ,
\end{equation}
which describes a harmonic oscillator with time-dependent frequency given by
\begin{equation}
\label{time-dependent-frequency}
\omega^2(t) = \omega_c^2 - \frac{{\ddot a}(t)}{a(t)}\, .
\end{equation}
In view of the isotropy, there is no loss of generality  in considering only one-dimensional motion in the $x^1$-direction. 

 The one-dimensional  Schr\"odinger equation associated with the above Lagrangian  is 
\begin{equation}
\label{Schrodinger-equation}
i\hbar \frac{\partial \Psi}{\partial t} = -\frac{\hbar^2}{2m}\frac{\partial^2 \Psi}{\partial x^2} + \frac{m\omega^2(t)}{2}\, x^2 \Psi\, ,
\end{equation}
where $x \equiv x^1$. It turns out that a  complete orthonormal set of solutions to Eq. (\ref{Schrodinger-equation})
is known in the form \cite{Khandekhar}
\begin{equation}
\label{orthonormal-solutions}
\psi_n(x,t)= \bigg(\frac{1}{2^n n!}\frac{\beta}{\sqrt{\pi}}\bigg)^{1/2}\exp \Bigl[ -i(n+ \frac{1}{2}) \gamma + \frac{im}{2\hbar}(i{\dot \gamma} + {\dot s}/s)x^2\Bigr] H_n(\beta x)\, ,
\end{equation}
where $H_n$ is the $n$-th Hermite polynomial, $\beta$ and $\gamma$ are functions of time satisfying
\begin{equation}
\label{beta-gamma}
\beta(t)= (m{\dot \gamma}(t)/\hbar)^{1/2}\, ,\,\,\,\,\,\,\,\,\,\, {\dot \gamma}s^2=1\, ,
\end{equation}
while $s(t)$ is a solution to the nonlinear differential equation
\begin{equation}
\label{s}
{\ddot s}- s^{-3} + \omega^2(t) s =0\, .
\end{equation}
Note that $\beta (t)$ and $\gamma (t)$ are immediately given as soon as $s(t)$ is found. It is shown in the Appendix that the above nonlinear equation for $s$ can be solved in terms of the linearly independent solutions of the classical equation of motion ${\ddot x} + \omega^2(t) x=0$.

For a  harmonic oscillator in the state $\psi_n$ it seems  reasonable to take the position dispersion
\begin{equation}
\label{size-definition}
(\Delta X)_n  = \sqrt{\langle \psi_n\vert x^2 \vert \psi_n\rangle  - \langle \psi_n\vert x \vert \psi_n\rangle^2}
\end{equation}
for its size. By noticing that the position probability density $\vert \psi_n (x,t)\vert^2$  takes exactly the same form as that for a standard harmonic oscillator of constant frequency $\omega$ if one  sets $\omega= \hbar \beta^2/m$, $(\Delta X)_n$ is readily calculated as \cite{Schiff}
 \begin{equation}
\label{size-calculated}
(\Delta X)_n (t)= \frac{\sqrt{n+ 1/2}}{\beta (t)} = \, \sqrt{\frac{(n+ 1/2)\hbar}{m}}\, s(t)\, ,
\end{equation}
where Eqs. (\ref{beta-gamma}) have been used. Thus, the oscillator's size is governed by the function $s(t)$. Therefore, strictly speaking, the cosmic expansion typically affects all bound systems, big or small.

Let us first consider the present de Sitter stage of exponential expansion dominated by the cosmological constant: $a(t)= a_0e^{H_0t}$. Then Eq. (\ref{time-dependent-frequency}) yields a constant frequency  $\omega_{eff} = \omega_c^2 - H_0^2$. Equation (\ref{s}) is solved by a constant $s=\omega_{eff}^{-1/2}$, and the wave functions (\ref{orthonormal-solutions}) reduce to the usual stationary states for an oscillator with constant frequency $\omega_{eff}$. Therefore, the oscillator does not expand at all, its size given by Eq. (\ref{size-calculated}) remains constant, in agreement with  Bonnor's result \cite{Bonnor}. This analysis is not valid if $\omega_c \leq H_0$,  in which case $\omega_{eff}  \leq 0$. Then the classical system is not bound and $x(t)$  grows either linearly or exponentially, resembling the all-or-nothing behavior found by Price and Romano \cite{Price}. The corresponding quantum system has no stationary bound states. The current accepted value of the Hubble time  is  $T_0=H_0^{-1} \approx 14$ Gyr. With $\omega_c=2\pi /T_c$, where $T_c$ is the system's characteristic time, $\omega_c \leq H_0$ requires $T_c > 80$ Gyr. The predicted growth is understandable, as one would hardly expect that a system with so long a characteristic time should be bound. 

Neutral atoms and cosmic  structures started to form  only after matter became dominant over radiation. Therefore,  in order to describe  the matter-dominated era before the cosmological constant takes over,  it  is appropriate to put  $a(t) = a_0t^{2/3}$, which yields $\omega^2(t) = \omega_c^2 + 2/9t^2$.  It is proved in the Appendix that Eq. (\ref{s}) is solved by 
\begin{equation}
\label{s-matter-dominated}
s(t) = \Bigl( \frac{\pi t}{2}\Bigr)^{1/2} \bigl[ J_{1/6}(\omega_c t)^2 + Y_{1/6}(\omega_c t)^2\bigr]^{1/2}\, ,
\end{equation}
where $J_{1/6}$ and $Y_{1/6}$ are Bessel functions of the first and second kind, respectively.
If $\omega_ct \gg 1/6$ the Bessel functions take the asymptotic forms \cite{Hildebrand}
\begin{equation}
\label{Bessel-asymptotic}
J_{1/6}(\omega_ct) \rightarrow \bigg( \frac{2}{\pi \omega_c t}  \bigg)^{1/2} \cos (\omega_c t-\pi/3)\, ,\,\,\,\,\,
Y_{1/6}(\omega_ct) \rightarrow \bigg( \frac{2}{\pi \omega_c t}  \bigg)^{1/2} \sin (\omega_c t-\pi/3)\, ,
\end{equation}
and it follows from (\ref{s-matter-dominated}) that
\begin{equation}
\label{s-asymptotic}
s(t)  \rightarrow \omega_c^{-1/2} = \mbox{constant} \, .
\end{equation}
This is just the asymptotic behaviour one must require of $s$, since $\omega (t) \to \omega_c $ as $t \to \infty$.

According to the cosmological model that best fits the current observational data \cite{Ryden}, from the time matter became dominant over radiation ($t_{rm} \approx 10^4$ yr) to the time when the cosmological constant began to dominate over matter ($t_{m\Lambda} \approx 10$ Gyr), the universe expanded by the factor 
 \begin{equation}
\label{expansion-factor}
\frac{a(t_{m\Lambda})}{a(t_{rm})} = \bigg( \frac{t_{m\Lambda}}{t_{rm}}\bigg)^{2/3} \approx 10^4  \, .
\end{equation}  
The natural frequency $\omega_c$ can be written as $\omega_c = 2\pi /T_c$, where $T_c$ is the system's  characteristic time. For a hydrogen atom, with characteristic time $T_h \approx 10^{-16}s$, we have $\omega_h t_{rm} \approx 10^{28}$. According to Eq. (\ref{s-asymptotic}), $s$ remains constant throughout and the atom does not expand. 
For  a system as large as our galaxy, the Sun's estimated period of revolution around the galactic center gives rise to the characteristic time $T_G \approx 200$ Myr. Even assuming the Milky Way was formed when the universe was only 500 million years old --- call this $t_f$ ---, at our galaxy's birth  one had  $\omega_G t_f \approx  2\pi \times .5\, \mbox{Gyr} /.2\, \mbox{Gyr} \approx 15$. For so large  a value of their argument, the Bessel functions of order $1/6$ have already reached their asymptotic forms, so that for all practical purposes $s=\mbox{constant}$ and the expansion of our galaxy has been negligible as compared to the overall expansion of the universe, which from $t_f \approx .5$ Gyr to $t_{m\Lambda} \approx 10$ Gyr stretches by a factor of about 7, as calculated from (\ref{expansion-factor}) with $t_{rm}$ replaced by $t_f$. This is confirmed by an explicit computation \cite{Casio} of the oscillator's stretching factor for $\omega_c = \omega_G$:
\begin{equation}
\label{galaxy-expansion-factor}
\frac{s(t_{m\Lambda})}{s(t_f)}\bigg\vert_G =  \sqrt{20}\, \frac{\bigl[ J_{1/6}(300)^2 + Y_{1/6}(300)^2\bigr]^{1/2}}
{\bigl[ J_{1/6}(15)^2 + Y_{1/6}(15)^2\bigr]^{1/2}} \approx 1.0001\, .
\end{equation}

For a  galaxy cluster, the estimate $T_{GC} \approx 10$ Gyr seems not too far off the mark. In this case, $\omega_{GC} t_f \approx  2\pi \times .5\, \mbox{ Gyr} / 10\, \mbox{Gyr} \approx 0.3$, whereas  $\omega_{GC} t_{m\Lambda} \approx  2\pi \times 10\, \mbox{Gyr}/ 10\, \mbox{Gyr} \approx 6$. From
Eqs. (\ref{size-calculated}) and (\ref{s-matter-dominated}) with $\omega_c= \omega_{GC}$ it follows that the cluster grows  by the factor \cite{Casio}
\begin{equation}
\label{cluster-expansion-factor}
\frac{s(t_{m\Lambda})}{s(t_f)}\bigg\vert_{GC} = \sqrt{20}\, \frac{\bigl[ J_{1/6}(6)^2 + Y_{1/6}(6)^2\bigr]^{1/2}}
{\bigl[ J_{1/6}(0.3)^2 + Y_{1/6}(0.3)^2\bigr]^{1/2}} \approx 1.13\, .
\end{equation}
Therefore, our model implies that during the entire  matter domination era a galaxy cluster gets  only thirteen percent bigger as compared to a factor 7 of overall expansion. 

\section{Conclusion}

Our model is so simple that it lends itself to a full and exact quantum treatment. This, in spite of the model's  crudeness,  permits a clear analysis of to what extent a bound system grows influenced by the overall cosmic expansion. It turns out that the degree of expansion of a bound  system appears to be fundamentally determined by a characteristic time scale associated with the system. Of course such a time scale is strongly correlated to the size of the bound structure, and  our elementary model indicates that even galaxy clusters essentially do not grow in response to the general Hubble flow.  

\section*{Appendix: Solving Eq. (\ref{s})}

{\bf Theorem.} Let $u$ and $v$ be two linearly independent solutions of the linear differential equation ${\ddot x} + \omega^2 (t) x=0$. Then
\begin{equation}
\label{solution-s}
s(t)= A [u(t)^2 + v(t)^2]^{1/2}
\end{equation}
is a solution to Eq. (\ref{s}) if the constant $A$ is chosen as $A = \vert W(u,v) \vert^{-1/2} $, where $W(u,v)=u{\dot v}-{\dot u}v$ is the Wronskian of $u$ and $v$. 

\noindent {\it Proof.} From (\ref{solution-s}) we get
\begin{equation}
{\dot s}= A \frac{u{\dot u} + v{\dot v}}{(u^2 + v^2)^{1/2}}\, , \,\,\,\,\,\,\, {\ddot s}= A \bigg[
\frac{u{\ddot u} + {\dot u}^2 + v{\ddot v} + {\dot v}^2 }{(u^2 + v^2)^{1/2}} -  \frac{(u{\dot u} + v{\dot v})^2}{(u^2 + v^2)^{3/2}}\bigg]\, ,
\end{equation}
so that
\begin{eqnarray}
s^3{\ddot s}& = & A^4\bigl[ (u^2 + v^2)(u{\ddot u} +{\dot u}^2 + v{\ddot v} + {\dot v}^2) 
- (u{\dot u} + v{\dot v})^2\bigr]\nonumber \\
& = & A^4\bigl[ u^3{\ddot u} + u^2 v {\ddot v} + u^2 {\dot v}^2 + v^2 u {\ddot u} +v^2 {\dot u}^2 + v^3{\ddot v}
- 2uv{\dot u}{\dot v}\bigr]\nonumber \\
& = & A^4\bigl[ -\omega^2 u^4  -2 \omega^2 u^2 v^2  + u^2 {\dot v}^2 + v^2{\dot u}^2-\omega^2 v^4 
- 2uv{\dot u}{\dot v}\bigr]
\, ,
\end{eqnarray}
where we have used ${\ddot u} = -\omega^2 u$ and ${\ddot v} = -\omega^2 v$. Therefore,
\begin{eqnarray}
\label{solucao}
s^3{\ddot s} + \omega^2 s^4 & = &  A^4\bigl[ -\omega^2 u^4  -2\omega^2 u^2 v^2  + u^2 {\dot v}^2 +  v^2{\dot u}^2   \nonumber \\
& - &  \omega^2 v^4 - 2uv{\dot u}{\dot v} + \omega^2(u^4+ 2u^2v^2+v^4)\bigl] \nonumber \\
& = & A^4\bigl[ u^2 {\dot v}^2  
- 2uv{\dot u}{\dot v} + v^2{\dot u}^2) = A^4 [u{\dot v} - {\dot u}v]^2 
\, .
\end{eqnarray}
With $\, W(u,v) = u{\dot v} - {\dot u}v\,$ we have
\begin{equation}
\frac{dW(u,v)}{dt} = u{\ddot v} + {\dot u}{\dot v} - {\ddot u}v - {\dot u}{\dot v}= -\omega^2 uv + \omega^2 uv =0 \,\, \Longrightarrow \,\, W(u,v) = \mbox{constant}\, .
\end{equation}
Furthermore, $\, W(u,v) \neq 0\,$ because, by hypothesis,  $u$ and $v$ are linearly independent. Thus, taking $\, A= \vert W(u,v) \vert^{-1/2}\,$ it follows from (\ref{solucao}) that (\ref{solution-s}) is a solution to Eq. (\ref{s}). The proof is complete.

\bigskip

In the case of $\omega^2(t) = \omega_c^2 + 2/9t^2$, two linearly independent solutions of ${\ddot x} + \omega^2(t)x=0$ are   $u(t) =(\omega_ct)^{1/2}J_{1/6}(\omega_c t)$ and $v(t) = (\omega_ct)^{1/2}Y_{1/6}(\omega_c t)$, where  $J_{1/6}$ and $Y_{1/6}$ are Bessel functions of the first and second kind, respectively \cite{Hildebrand}. From the asymptotic forms (\ref{Bessel-asymptotic}) of the  Bessel functions $J_{1/6}$ and $Y_{1/6}$ it follows easily that  $W(u,v) = 2\omega_c /\pi $. Thus, the function $s$ given by (\ref{solution-s}) with $A= (2\omega_c /\pi)^{-1/2}$ takes exactly the form  (\ref{s-matter-dominated}).


\end{document}